\documentclass[prl,lengthcheck,superscriptaddress,showpacs,letterpaper,nofootinbib]{revtex4-1}

\usepackage{color}
\usepackage{graphicx}  
\usepackage{float}
\usepackage{amsmath}
\usepackage{subfigure}

\begin{document}

\title{Depressing de Sitter in the Frozen Future}

\author{Andrew P. Lundgren} \email{andrew.lundgren@LIGO.org} \affiliation{Albert-Einstein-Institut, Callinstr. 38, 30167 Hannover, Germany}
\author{Mihai Bondarescu} \email{mihai7@gmail.com} \affiliation{University of Mississippi, Oxford, MS, USA}
\author{Ruxandra Bondarescu} \email{ruxandra@physik.uzh.ch} \affiliation{Institute for Theoretical Physics, University of Zurich, Switzerland}

\begin{abstract}
In this paper we focus on the gravitational thermodynamics of the far future. Cosmological observations suggest that most matter will be diluted away by the cosmological expansion, with the rest collapsing into supermassive black holes. The likely future state of our local universe is a supermassive black hole slowly evaporating in an empty universe dominated by a positive cosmological constant. We describe some overlooked features of how the cosmological horizon responds to the black hole evaporation. The presence of a black hole depresses the entropy of the cosmological horizon by an amount proportional to the geometric mean of the entropies of the black hole and cosmological horizons. As the black hole evaporates and loses its mass in the process, the total entropy increases obeying the second law of thermodynamics. The entropy is produced by the heat from the black hole flowing across the extremely cold cosmological horizon. Once the evaporation is complete, the universe becomes empty de Sitter space that (in the presence of a true cosmological constant) is the maximum entropy thermodynamic equilibrium state. We propose that flat Minkowski space is an improper limit of this process which obscures the thermodynamics. The cosmological constant should be regarded not only as an energy scale, but also as a scale for the maximum entropy of a universe. In this context, flat Minkowski space is indistinguishable from de Sitter with extremely small cosmological constant, yielding a divergent entropy. This introduces an unregulated infinity in black hole thermodynamics calculations, giving possibly misleading results.
\end{abstract}

\maketitle

{\it Introduction.}
Our universe appears to have a small positive cosmological constant \cite{1998AJ....116.1009R,1999ApJ...517..565P,2011ApJS..192...18K}. Although this has had little effect on the past evolution of the universe, it will completely dominate the cosmological future \cite{2000ApJ...531...22K}. Matter will stream outwards through the cosmological horizon, leaving behind only the gravitationally bound part of our local universe. In due time, nothing will be left but supermassive black holes, slowly evaporating by Hawking radiation \cite{1974Natur.248...30H}. The future will asymptote to de Sitter space, the spacetime of a universe containing nothing but a positive cosmological constant.

This bleak future is actually very interesting from the perspective of gravitational thermodynamics. The cosmological horizon absorbs the matter and radiation that streams through it, and grows in response. The entropy of the horizon is given by its area \cite{1973PhRvD...7.2333B,1976PhRvD..13..191H} , so the cosmological horizon is producing entropy by absorbing the contents of the universe. If the cosmological constant is a true constant, then de Sitter spacetime is the final equilibrium state of our universe. The emptying of the universe is the approach to this equilibrium state.

We will model this process by the slow evaporation of a single non-spinning supermassive black in a de Sitter background. We will present some interesting features that we believe have been overlooked. In particular, we argue that de Sitter is a better background than the flat Minkowski spacetime for understanding gravitational thermodynamics. Minkowski space is typically considered to have zero entropy, but we take the alternative view that Minkowski is a limit of de Sitter as the cosmological constant goes to zero. In this limit, the entropy of the spacetime is infinite, which introduces an unregulated infinity into the calculation. 

In the de Sitter background, the gravitational effect of the black hole shrinks the area of the cosmological horizon. The fractional change in the area is small because the black hole horizon is small compared to the cosmological horizon. But the change in the area, and therefore the entropy, of the cosmological horizon is very large compared to the entropy of the black hole. Also, the Hawking temperature of the cosmological horizon is very small even compared to the black hole temperature, leading to the large production of entropy as heat flows across it. The presence of a black hole depresses the entropy of de Sitter, but the entropy rises as the black hole evaporates into the freezing cosmological horizon surrounding it. The thermodynamics of black hole evaporation is less clear in flat Minkowski spacetime, where the decrease in the gravitational entropy of the black hole forever remains in the outgoing Hawking radiation. In de Sitter, the increase in entropy can be calculated directly from the horizon areas, or seen as the result of heat flowing across the temperature difference between the two horizons.

{\it Temperature and Entropy in de Sitter.}
The existence of a cosmological constant is the simplest explanation for the accelerating expansion of the universe. In the past, the density of radiation and matter have dominated that of the cosmological constant, but the expansion of the universe dilutes the densities of radiation and matter while not affecting the cosmological constant. The cosmological constant will eventually completely dominate the density of the universe on cosmological scales, leading to an exponential expansion and a cold, nearly empty universe. The cosmological metric describing an empty universe with a cosmological constant is
\begin{equation}
ds^2 = - dt^2 + a^2(t) \left( dr^2 + ~ r^2 (d\theta^2 + \sin^2 \theta ~d\phi^2) \right)
\end{equation}
with scale factor $a(t) = e^{H t}$ and $H = \sqrt{\Lambda / 3}$, where $H$ is the Hubble parameter and $\Lambda$ is the cosmological constant. The metric can also be written in static coordinates as
\begin{eqnarray}
ds^2 &=& - \left(1 - \frac{\Lambda r^2}{3}\right) dt^2 + \left(1 - \frac{\Lambda r^2}{3}\right)^{-1}  dr^2 \nonumber \\ 
& & + ~ r^2 (d\theta^2 + \sin^2 \theta ~d\phi^2) ~.
\end{eqnarray}
We will refer to $r$ as the radius, although it is more properly referred to as the areal radius; it does not measure distance from the origin, but the area of a sphere concentric with the origin is $4 \pi r^2$.

In the static slicing, it is clear that there is a horizon at $r = \sqrt{3 / \Lambda}$. This has a thermodynamic interpretation; a horizon has entropy given by one-quarter the horizon area. For the de Sitter cosmological horizon,
\begin{equation}
S = \frac{3 \pi}{\Lambda} ~.
\end{equation}
Likewise, the temperature is related to the surface gravity of the horizon. There is a normalization issue that we will consider later, but for now we quote the standard result \cite{1977PhRvD..15.2752G}
\begin{equation}
T = \frac{1}{2 \pi} \sqrt{\frac{\Lambda}{3}} ~.
\end{equation}
Throughout this paper, we will use units where $G = \hbar = c = k_B = 1$, but we will give an example in physical units.

{\it The Flat Space Limit.}
Consider the limit as $\Lambda$ goes to zero. The metric becomes the Minkowski metric of flat spacetime. On physical grounds, flat spacetime is nearly indistinguishable from one with a very small cosmological constant. The entropy and temperature of Minkowski spacetime should then be those of de Sitter in the limit of very large horizon size. In this limit, the temperature is zero, which is not surprising. However, in the same limit the entropy diverges. This is very different from the typical assumption that because Minkowski space does not have a horizon, it has zero entropy.

It is possible that instead of a true cosmological constant, there is instead a slow-rolling scalar field \cite{PhysRevLett.75.2077, PhysRevLett.82.896}, whose potential energy has the same effect 
as $\Lambda$. When the field rolls to a lower potential energy, the effective $\Lambda$ decreases and the entropy increases. Our expectation that 
scalar fields can roll down their potentials but not up corresponds to the expectation that entropy can increase but not decrease. Cosmic acceleration in this case is attributed to the fact that the universe has not reached its true vacuum state for dynamical reasons. If the true cosmological constant is zero, the end state can be Minkowski space. 

If Minkowski space has divergent entropy, then it is not a good background in which to study black hole thermodynamics. Such calculations would have an unregulated infinity in them, giving possibly misleading results. We will now show that the evaporation of a Schwarzschild black hole in de Sitter space has a satisfying thermodynamic interpretation and is quite different from the picture in Minkowski space.

{\it Black Hole Evaporation.}
We now consider a non-rotating black hole in a universe with a positive cosmological constant \cite{1977PhRvD..15.2738G, PhysRevD.34.1700}. The de Sitter-Schwarzschild metric is 
\begin{equation}
ds^2 = - f(r) dt^2 + f(r)^{-1} dr^2 + r^2 ( d\theta^2 + \sin^2 \theta d\phi^2 )
\end{equation}
where
\begin{equation}
f(r) = 1 - \frac{2 M}{r} - \frac{r^2}{\alpha^2} ~.
\end{equation}
For convenience, we use $\alpha = \sqrt{3 / \Lambda}$ to parameterize the cosmological constant; keep in mind that $\alpha$ grows with decreasing $\Lambda$. Many of the symmetries of de Sitter spacetime are broken by the presence of the black hole, but there is still a time symmetry which is explicit in the static form of the metric.

In addition to the cosmological horizon, there is now a black hole horizon. Each is a sphere where $f(r) = 0$. There is a third, negative root which we discard. The horizons are roughly at $r_{BH} \approx 2 M$ and $r_{C} \approx \alpha$. Making the approximation that $M \ll \alpha$, the radii can be expanded as
\begin{eqnarray}
r_{BH} &\approx& 2 M + 8 \frac{M^3}{\alpha^2} + 96 \frac{M^5}{\alpha^4} + \ldots \\ 
r_{C} &\approx& \alpha - M - \frac{3}{2} \frac{M^2}{\alpha} - 4  \frac{M^3}{\alpha^2} + \ldots ~. 
\end{eqnarray}
We can now find the entropies associated with each horizon; each is one-quarter the horizon area. These are
\begin{eqnarray}
S_{BH} &\approx& \pi ( 4 M^2 + 32 \frac{M^4}{\alpha^2}) + \ldots \\
S_{C} &\approx& \pi ( \alpha^2 - 2 \alpha M - 2 M^2 - 5 \frac{M^3}{\alpha} ) + \ldots \\
S_{tot} &\approx& \pi ( \alpha^2 - 2 \alpha M + 2 M^2 - 5 \frac{M^3}{\alpha} ) + \ldots
\end{eqnarray}
The important feature is that the total entropy $S_{tot}$ \emph{increases} as the mass of the black hole decreases, because the $- 2 \pi \alpha M$ term in the cosmological horizon entropy is much larger than $S_{BH}$. We know that a black hole will emit Hawking radiation and lose mass in the process, eventually evaporating away to nothing. We see that the total entropy will increase in this process, as required by the second law of thermodynamics. We will now consider the thermodynamics in more detail.

The temperature of each horizon is $T = \frac{\kappa}{2 \pi}$. The surface gravity $\kappa$ is defined by
\begin{equation} \label{EqSurfaceGravity}
k^\alpha \nabla_\mu k^\beta = \kappa ~ k^\beta  ~,
\end{equation}
where $k$ is the timelike Killing vector that expresses the time-invariance of the metric. We write
\begin{equation}
k = \mathcal{N} \frac{\partial}{\partial t}
\end{equation}
where $\mathcal{N}$ is a normalization constant. The static coordinates we are using have problems at the horizons, so we must do the calculation in another coordinate system such as Painlev\'e-Gullstrand; the result is
\begin{equation}
\kappa =  \mathcal{N} \frac{f'(r)}{2} = \mathcal{N} \left( \frac{M}{r^2} - \frac{r}{\alpha^2} \right) ~.
\end{equation}
The normalization is fixed by requiring that $k_\alpha k^\alpha = -1$ at the observer who measures the temperature \cite{1977PhRvD..15.2738G, lrr-2001-6}, so that.
\begin{equation}
\mathcal{N} = \frac{1}{\sqrt{-g_{t t}}} = \frac{1}{\sqrt{f(r)}} ~.
\end{equation}
The normalization accounts for the gravitational redshift when comparing the temperatures measured by observers at different radii. In a Minkowski background, $\mathcal{N} = 1$ at asymptotic infinity. The temperatures are
\begin{eqnarray}
T_{C}(r) &=& \frac{\kappa_{C}}{2 \pi} \\
&\approx& \frac{1}{2 \pi \alpha \sqrt{f(r)}} \left( 1 - 2 \frac{M}{\alpha} - \frac{7}{2} \frac{M^2}{\alpha^2} + \ldots \right) \nonumber \\
T_{BH}(r) &=& - \frac{\kappa_{BH}}{2 \pi} \\
&\approx& \frac{1}{8 \pi M \sqrt{f(r)}} \left( 1 - 16 \frac{M^2}{\alpha^2} + \ldots \right) \nonumber ~.
\end{eqnarray}
We must reverse the sign of the surface gravity on the cosmological horizon since it is an outer horizon rather than an inner horizon. Note that the surface gravity is evaluated at the relevant horizon, while the redshift is evaluated at the radius of interest. These are the temperatures as measured by static observers, i.e. those that flow along the Killing vector $k^\mu$ so they remain at a constant value of $r$.

The black hole always has a higher temperature than the cosmological horizon unless the black hole grows to a mass of $M = \frac{1}{3 \sqrt{3}} \alpha$, at which point the temperatures are equal but the two horizons merge. We clearly see that this is a non-equilibrium situation because there are two non-equal temperatures. Heat will flow from hot to cold, causing the black hole to evaporate and producing entropy.

We separately consider the changes in entropy of the two horizons. We use the thermodynamic expression
\begin{equation} \label{EqEntropyProduction}
\delta S = \frac{\delta Q}{T}
\end{equation}
relating the change in entropy $\delta S$ to the heat absorbed or emitted by the horizon $\delta Q$. We model the evaporation of the black hole as a small change in $M$, with $\alpha$ of course remaining fixed. The luminosity of the black hole is small enough that this quasi-static approximation is reasonable. For a small change $\delta M$ in the black hole mass, the heat emitted is
\begin{equation}
\delta Q(r) = \frac{\delta M}{\sqrt{- g_{t t}}} \label{EqRedshiftedHeat} ~,
\end{equation}\
where $(- g_{t t})^{-1/2}$ is a redshift factor to give the heat measured by an observer at radius $r$.

Substituting Eq. \eqref{EqRedshiftedHeat} into Eq. \eqref{EqEntropyProduction} gives
\begin{equation}
\frac{\partial S}{\partial M} \delta M = \frac{1}{T(r)} \frac{\delta M}{\sqrt{- g_{t t}}}  ~. \label{EqChangeInEntropy}
\end{equation}

It is simple to evaluate the left and right-hand sides of Eq.~\eqref{EqChangeInEntropy} for either the black hole or cosmological horizon, although we omit the algebra. For each horizon, the change in entropy is exactly what is expected from the heat crossing the horizon, $\delta S = T \delta Q$. This holds even without the approximation that $M \ll \alpha$, and also holds regardless of the value of $r$ where it is evaluated, since the redshift factors $(-g_{t t})^{-1/2}$ in the heat and the temperature cancel. The black hole horizon loses a small amount of entropy by emitting some heat at a relatively high temperature. The cosmological horizon gains a much larger amount of entropy by absorbing the heat at a lower temperature. It is significant that the normalization constant is determined using only one horizon, which then correctly predicts the entropy production at the other horizon.

We clearly see that black hole evaporation proceeds in the de Sitter background because it leads to an increase in entropy. Heat flows from hot to cold, and the entropy production at each horizon is exactly what is expected from Eq.~\eqref{EqEntropyProduction}. This is clearer than the corresponding  situation in Minkowski space, where the entropy lost by the black hole is forever left in the form of Hawking radiation. There is of course an intermediate state even in de Sitter where the Hawking radiation is traveling outward toward the cosmological horizon, but this is a short time compared to the evaporation timescale of a supermassive black hole in our universe.

As an example, we can imagine that the Local Supercluster collapses to form a supermassive black hole \cite{2000ApJ...531...22K}. The mass of this black hole will be roughly $10^{15} M_\odot$ \cite{2007A&A...476..697E}, giving a Schwarzschild radius of $300$ light years. The cosmological horizon has a radius of $18$ billion light years. The entropies are measured in units of Planck area $A_{Pl} = G \hbar c^{-3} \approx 3 \times 10^{-102}$ square light years. The temperature is converted from an inverse distance with a factor $\hbar c \approx 2 \times 10^{-23}$ eV light years. In these units, the black hole horizon has an entropy of $10^{107}$ and a temperature\footnote{These are the temperatures without the $g_{t t}^{-1/2}$ redshift factor included.} of $10^{-27}$ eV, while the cosmological horizon has an entropy  $10^{122}$ and a temperature of $10^{-34}$ eV. The entropy produced by the evaporation of the black hole is $10^{114}$, which is clearly much larger than the entropy of the black hole horizon itself (it is nearly the geometric mean of the two horizon entropies).

{\it Weyl Curvature Hypothesis.}
We can now make contact with the Weyl Curvature Hypothesis of Penrose \cite{1979grae.book.....H}, which states that the difference in entropy between the initial and final states of the universe is related to the growth of the Weyl curvature. The beginning of the universe is likely to have very small Weyl curvature but it grows due to the production of singularities in the late universe. However, the de Sitter metric has zero Weyl curvature but extremely large entropy, so it seems there cannot be a direct relation between Weyl curvature and gravitational entropy. We conjecture that there is a more subtle relation between the Weyl curvature and the gravitational entropy. The Einstein equation relates the Ricci curvature to the cosmological constant and local stress-energy tensor $T_{\mu \nu}$ by
\begin{equation}
R_{\mu \nu} = 8 \pi G (T_{\mu \nu} - \frac{1}{2} g_{\mu \nu} T) + \Lambda g_{\mu \nu} ~.
\end{equation}
This does not determine the full Riemann curvature of the spacetime; the Weyl tensor is free to evolve except for the Bianchi identities, which effectively provide boundary conditions relating Weyl to Ricci and therefore to $T_{\mu \nu}$. The Ricci curvature then roughly plays the role of macroscopic observables, while the entropy comes from coarse-graining over states of the Weyl tensor. For instance, gravitational waves are modes of the Weyl tensor. The effect of the cosmological constant is to not allow gravitational wave modes with very long wavelengths. When a black hole evaporates, the cosmological horizon grows, and it could be that this allows more gravitational wave modes to exist and so increases the number of accessible microstates in the Weyl tensor. This idea needs further development.

{\it Discussion and Conclusions.}
For a fixed value of the cosmological constant, the maximum entropy of the Schwarzschild-de Sitter solution is obtained in empty de Sitter space. The presence of a black hole shrinks the cosmological horizon and depresses the total entropy. Both the black hole and cosmological horizons emit Hawking radiation, but the black hole is hotter so the net flow of energy is outward from the black hole towards the cosmological horizon. The radiation passes through the cosmological horizon, carrying a small amount of entropy away from the black hole. When the radiation crosses the cosmological horizon, the entropy in the radiation disappears from the perspective of an observer between the two horizons. However, by Eq. \eqref{EqEntropyProduction}, it is clear that this heat flowing across the much colder cosmological horizon produces much more entropy than was lost by the black hole.

An interesting feature is that the black hole can be replaced by a spherical star with the same mass. The metric external to the star will be the same, so the cosmological horizon will behave exactly the same. There will be no black hole horizon, but if the star loses mass in the form of radiation it will produce entropy as the radiation flows across the cosmological horizon. This suggests that any matter within the horizon depresses the entropy relative to empty de Sitter. Also, the collapse of a massive star to a black hole does increase the total entropy, since a black hole horizon forms (and has larger entropy than the star) but the cosmological horizon is not affected.

Black hole evaporation is a non-equilibrium process, so it produces entropy; the end of the evaporation should correspond to maximum entropy and to the equilibrium state. There is evidence \cite{1988CQGra...5.1349D, 2003CQGra..20.2753D} that empty de Sitter space has the maximum entropy of spacetimes with a fixed cosmological constant. The equilibrium state should in a sense be the most generic state. The equilibrium state of a gas in a box is one where the gas is uniformly distributed and featureless. de Sitter space is maximally symmetric, meaning there is a full set of ten symmetries expressed by Killing vectors (one time and three space translations, three rotations, and three boosts). Minkowski space possesses the same symmetries but does not have the preferred scale that the cosmological constant provides, so it is even more symmetric. The cosmological constant could equally well be regarded as an energy scale or as setting the maximum total entropy of our universe.

If the true cosmological constant is zero, then the maximum entropy state of the universe is Minkowski. The cosmological constant behavior could then be mimicked by a quintessence scalar field \cite{PhysRevLett.75.2077, PhysRevLett.82.896} with an ultralight mass of $\sim 10^{-33}$ eV that slowly rolls down its potential. Another ultralight scalar field of mass Ê$\sim 10^{-23} eV$ is a possible model for dark matter \cite{2000PhRvL..85.1158H, 2010ApJ...715L..35L}. This scalar field would have already reached the bottom of its potential and formed dark matter particles that Bose-condensed into the halos around galaxies. Once the dark energy scalar field reaches the final minima of its potential, it could form new particles as well converting a small fraction of the cosmological constant entropy into entropy for the particles. The temperature and entropy of both the dark matter and the quintessence scalar field particles would be interesting from a thermodynamics perspective. In addition, black holes are ineffective dark matter accretors due to the low viscosity of dark matter particles and it is unclear how scalar field entropy would flow into the cosmological horizon. Modeling such a scenario is subject to future work.

{\it Acknowledgements.}  RB is supported by the Dr. Tomalla Foundation and the Swiss National Foundation. We thank Badri Krishnan, Alex Nielsen, and Bjoern Schmekel for useful discussions and Phillipe Jetzer for encouragement and support. APL thanks Alex Nielsen for pointing out that the black hole can be replaced by a spherical star. MB thanks Edward Witten for useful discussions, and APL is especially grateful to James York Jr. for introducing him to black hole thermodynamics.

\end{document}